# Mathematical Singularities in the Farthest Confines of the Universe—And a Brief Report on Its Evolutionary History

Emilio Elizalde

Consejo Superior de Investigaciones Científicas, Institute of Space Sciences ICE-CSIC and Institut d'Estudis Espacials de Catalunya, IEEC, Campus UAB, Carrer de Can Magrans s/n, 08193 Cerdanyola del Vallès (Barcelona), Spain; elizalde@ice.csic.es

**Abstract:** It is advisable to avoid and, even better, demystify such grandiose terms as "infinity" or "singularity" in the description of the cosmos. Its proliferation does not positively contribute to the understanding of key concepts that are essential for an updated account of its origin and evolutionary history. It will be here argued that, as a matter of fact, there are no infinities in physics, in the real world: all that appears, in any given formulation of nature by means of mathematical equations, actually arises from extrapolations, which are made beyond the bounds of validity of the equations themselves. Such a crucial point is rather well known, but too often forgotten, and is discussed in this paper with several examples; namely, the famous Big Bang singularity and others, which appeared before in classical mechanics and electrodynamics, and notably in the quantization of field theories. A brief description of the Universe's history and evolution follows. Special emphasis is put on what is presently known, from detailed observations of the cosmos and, complementarily, from advanced experiments of very high-energy physics. To conclude, a future perspective on how this knowledge might soon improve is given.

**Keywords:** singularity; infinite; Big Bang; universe evolution; scientific theory

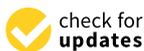

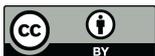

## 1. Introduction: Fleeting Biblical Reference

The first book of the Bible, the Book of Genesis, contains an account of the creation of the world. We read there that God created the world in seven days. This narrative, although precise in some details, is quite schematic. It conveys the idea that, at the beginning of time and starting from nothing, God made the Universe we now have (Figure 1). However, rather than from nothing, according to John 1:1 in the New Testament, it could have been from an "energy": the "Word"—which John says was with God and was God himself.

Technically speaking, this is a fairly static and somewhat mechanistic vision, although on a very grand scale. One should note that, eventually, the whole work was completely finished. The evolution of the cosmos is not sufficiently considered. It should be observed, too, that the biblical creation of the world also contains another crucial stage, namely the creation of Man: from the dust of the ground, God breathed into him the breath of life and consciousness, and then into Eve, to become partners. This paper will not deal at all with the last issue, which is of enormous complexity, and is today being investigated by other means.

This very brief introduction should be taken as a mere historical and comparative reference, without further ado. It is not the purpose of the present work to delve into the biblical description, nor to criticize it. Some consider it as a poetic story, inherited from others coming from older cultures; but it is certainly fascinating and has some points of contact with the much more accurate view that science gives us today. It is the latter that will be considered below, in an affordable way, but, at the same time, with all the rigor that, what has been so far discovered and experimentally verified, provides.

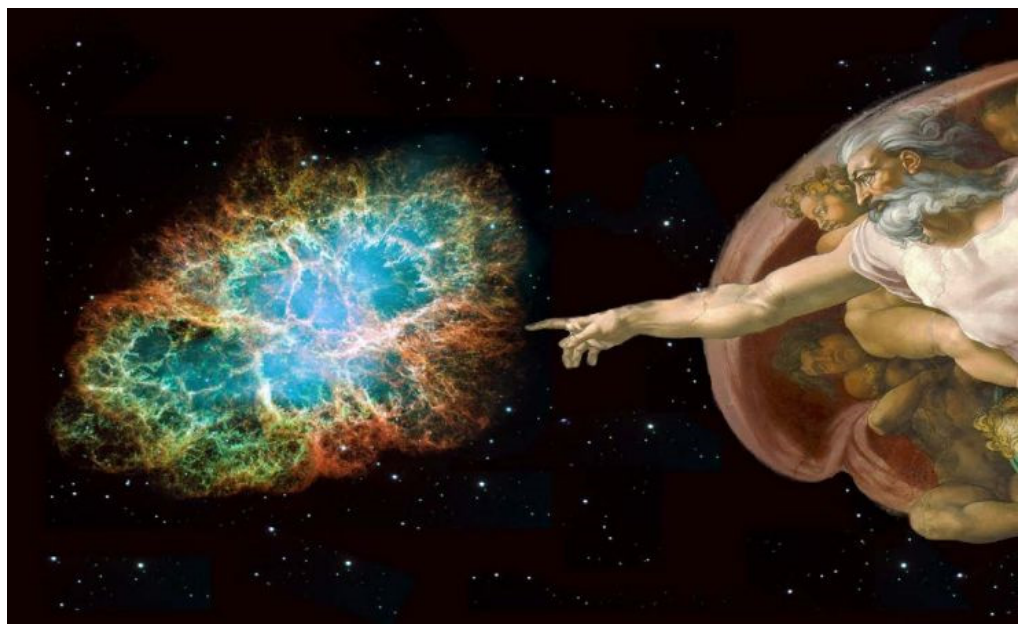

**Figure 1.** Artistic representation of God's creation of the Universe.

At the risk of simplifying the comparison excessively, one could already advance, as an appetizer, that it might look as if, in the modern descriptions of the origin of everything, the latent (and, for many, explicit) mystery in the version of Genesis has been replaced nowadays by another equally unfathomable one, which is however, for our rational intellect, much more precise, well posed, and apparently rigorous: The Big Bang singularity. The paper will deal with this singularity, as well as with others that so often arise in the mathematical equations corresponding to physical theories of different kinds.

What will follow is a transcription—though substantially expanded in several aspects—of the content of two recent talks given by the author: one of them in a removable booth in the middle of the vivid Rambla del Raval, in Barcelona, within the framework of the Barcelona Science Festival, and the other at the stylish Auditorium Barradas of Hospitalet de Llobregat, also in Spain, on the occasion of the European Researcher's Night 2022, respectively. The thorough preparation of these talks led the author to reflect in depth on some of the points that will follow. We should praise, by the way, such great initiatives, which seek to bring top scientific subjects to ordinary people in a solvent way, far from the cartoonish simplification, and to awaken the interest and passion for true science, which we miss so much in our society.

The structure of the paper is as follows. In Section 2 the Big Bang singularity and other relevant infinities are considered. Some key issues are discussed in several subsections devoted, respectively, to the limited range of validity of scientific theories, the definition of a mathematical function, and the ways infinities are dealt with, by means of regularization and renormalization processes. In fact, only a brief, historical account of some of them will be provided, with the purpose of stressing the main point here, which is namely that there are no infinities in nature and that they only arise because the mathematical equations describing the corresponding physical theory are extended beyond their actual domain of validity. In Section 3, the first stages of the origin and evolution of our universe are discussed, up to the reheating epoch. Emphasis is put in the fact that they involve deep knowledge of particle physics at very high energies, indeed, the highest ones that are presently attainable in particle accelerators. In Section 4, a brief discussion of the recombination and reionization epochs of the Universe's evolution is given. Some current and future missions for the exploration of the cosmos will be presented in Section 5. In Section 6, a general discussion of the main issues considered in the paper is presented. Finally, Section 7 contains some closing conclusions.

## 2. The Big Bang Singularity and Other Relevant Divergences

The traditional theory of the Big Bang uses the Theory of General Relativity (GR) to describe the Universe and, in particular, its origin; that is, the origin of everything, of space and time, of matter and energy, in a word, of all that we see. In the current, most commonly accepted description, everything begins, somehow mysteriously one must add, in the so-called Big Bang singularity. However, the first (and serious) problem is that this extraordinary theory, for many, the most beautiful theory ever created (there are no adjectives to qualify it as it deserves), turns out to be not fitted for studying the Universe when it was very small, let alone its origin.

Let us now reflect for a while, as the image on the right side of the picture (Figure 2) invites us to do. At a much more comprehensible level—setting our feet on the ground and without having to travel that far back in time—just consider the atomic level, which is much more familiar and closer to our understanding. Every physicist knows, since at least a whole century now, that GR is of no use to describe the behavior of atoms, let alone of their nuclei (several orders of magnitude smaller). However, all of them are nowadays ordinary objects, on which much of our present scientific and technological developments are based. However, there, at the atomic and nuclear levels, the Einstein equations fail with exactly the same crash as Newtonian mechanics does; since, on those scales, GR does not add anything new to the Newtonian results! At the very least we physicists (and scientists of any sort I would dare to say) should be fully aware of this circumstance; although sometimes, perhaps enchanted by the extreme beauty of the Einsteinian theory, it does seem as if we forgot.

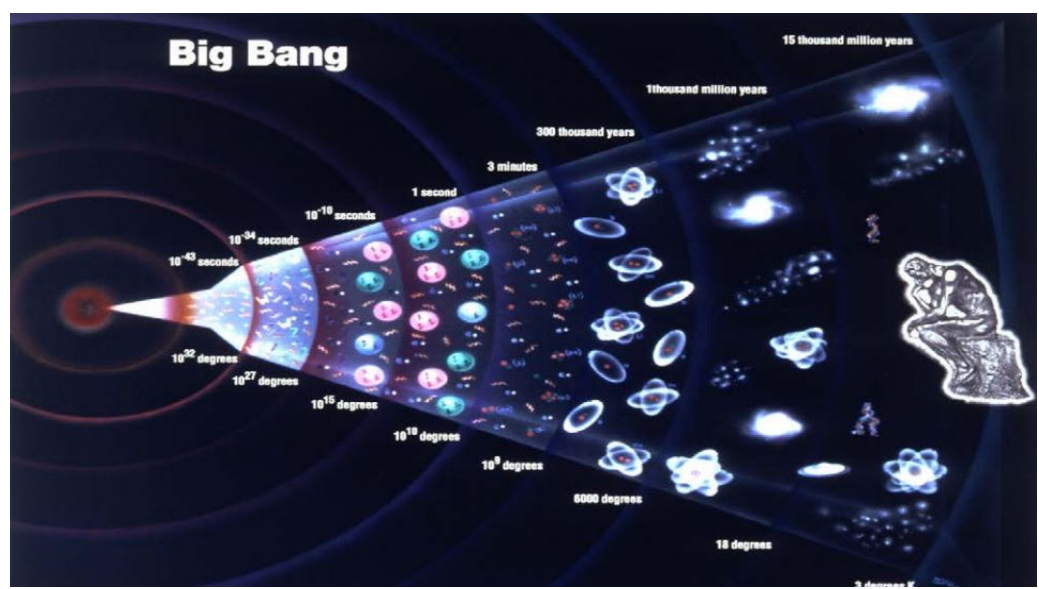

**Figure 2.** Schematic representation of the Big Bang singularity and subsequent universe evolution. Credit: grandunificationtheory.com.

And now, just reflect a bit more. The consequence of the above is that all those ramblings, which, with the intention of impressing, are so often written about the "initial singularity", about a "point-like" primitive universe of "infinite density and temperature", and many other nonsense statements such as those —which even manage to sneak into some versions of Wikipedia (fortunately, not in all of them)—are all mere mathematical extrapolations without the slightest *physical* sense. We could continue arguing about this in many ways, apart from the reasons already given.

*2.1. The Limited Range of Validity of Scientific Theories*

The following message must stay clear: in the real world, in nature, there are *no* infinities! All those that have appeared, on so many occasions, in different theories and circumstances, all those that arise from the mathematical formulations of physical reality when attempting to describe it —such as the Newtonian laws, Einstein's field equations, or Maxwell's equations of electrodynamics—are, without exception, mathematical artifacts! They only pop up because it so happens that the equations of physics are always, when we are outside certain domains or specific conditions, not precise enough or even completely inappropriate to describe the real world.

A key issue to be understood here is that the equations that express, in mathematical language, the principles of a physical theory, whatever this may be, *always* have a *limited* range of validity. When one leaves such a domain, the equations start to become imprecise, and then stop being useful at all.

A very clear example of this issue are the Newtonian equations. For several centuries they were believed to be universal, literally: the laws of universal gravitation were true under *any* circumstances or conditions, until it became dramatically clear that this was not at all the case; rather, they had to be drastically corrected, both at small distances (quantum physics) and at high speeds (theory of relativity). However, it is also a fact that these brand-new theories have, in turn, their own *limits*, their respective ranges of validity: they are not universal either. In no way are they the final theories to describe our universe; at most, they are really good, complementary approximations to such.

*2.2. Mathematics for Comparison: The So Precise Definition of a Mathematical Function, and the Unreasonable Effectiveness of Mathematics in the Physical World*

Mathematicians, in the realm of their axiomatic world, parallel, in some ways, to the physical one, have a perfect understanding of this point, which is already embedded in the very definition of a concept as basic as that of a "mathematical function". As is taught in high-school or freshman courses, a mathematical function is not merely a "correspondence", *f* (that is, a mathematical expression or formula), but it also includes its "domain", *A*, and "range", *B*, that is, the initial and final sets on which the correspondence is defined and it is strictly valid, that is, where it actually makes sense. A mathematical function is defined, therefore, as the triple (*A*,*B*,*f*)—and never just as *f*. Note the absolute parallelism with what precedes, on the domain of validity of a physical theory [1,2].

A recurrent issue that has puzzled many renowned scientists and engineers over time, is that of the unreasonable effectiveness of mathematics in the physical world. Of the several references that are immediately brought to mind (Einstein himself expressed this idea on several occasions), I have selected two, namely the famous paper by Eugene P. Wigner entitled "The Unreasonable Effectiveness of Mathematics in the Natural Sciences" [3], where this issue is magnificently explained, and another one by Richard W. Hamming, with the shorter title "The Unreasonable Effectiveness of Mathematics", where the subject is treated from a complementary viewpoint and further examples are given, coming from engineering and other scientific disciplines [4]. Little will be added here to what can be found there but, in relation with the basic discussion, the following remark is in order.

It needs to be recalled that, in addition to classical mechanics, we owe to Isaac Newton another invention of extreme importance, which he did in parallel with Gottfried Wilhelm Leibniz, namely, infinitesimal calculus [5,6]. Many of the equations in science are differential ones and integration is performed according to the rules of calculus, too. Infinitesimal calculus is the mathematical study of continuous change, and its starting scenario is a rigorous, perfectly well-established definition of the mathematical continuum. This concept corresponds to the dense distribution of real numbers on the real line, then extended to two, three, and to any number of dimensions (even beyond those of the physical space), by Bernhard Riemann. In particular, the construction, step by step, of the real line, is one of the most exciting lessons of any university course on infinitesimal calculus. No student of physics or engineering should ever miss those lectures.

When attending, later, the lectures on atomic physics, it becomes suddenly clear how far apart are the two understandings: the one of the lower and lower levels of physical reality and that of the mathematical conception of the same dimensions. They have nothing to do with each other! And the mystery of the extraordinary effectiveness of mathematics in modeling physics grows and grows at the same pace.

Further to that, and looking again through a powerful microscope, one also realizes all of a sudden that already the simple definition of a mathematical "point" or "dot" is an abstraction of reality; and the same is true for those of a line, a plane, etc. There are *no* physical objects which do correspond to such mathematical, purely axiomatic constructions, in physics. Think, e.g., that graphene sheets—the thinnest to have been constructed ever— are at least one atom thick (which is not much, but still quite far from zero-depth). What would be our physical candidate for a mathematical dot? Of course, not a quantum dot, a quark perhaps? Maybe a neutrino, which is much smaller, but still rather big, as compared to zero-diameter. In s few words, a point-particle has no real sense. When confronted with this, a physicist should always ask something like: How small do you mean your dot is? Does it have a diameter of $10^{-10}$ cm? of $10^{-35}$ cm? of $10^{-120}$ cm? Likewise, the term "infinity" is also devoid of real sense. When one hears the words "infinite temperature" at the moment of the Big Bang, the immediate question should follow: What do you mean? Higher than $10^{50}$ K, or something? It is still a long, a very long way, to infinity, or to zero, in both these examples. The actually strange thing–a great puzzle—is that working consistently with those abstract mathematical entities, one does obtain results that are extremely close to the real results in the laboratory.

Whatever the case, and to summarize, what is undoubtedly wrong and inappropriate is to talk of point physical particles, infinite matter or charge densities, infinite temperatures, and the like—in short, divergences galore—when there are none. It is my point that even quick, vague, or cartoonish descriptions of reality should avoid these expressions, which simply make no sense and do not help to understand the actual physical concepts.

*2.3. Taming Infinities: Cantor, Euler, Riemann, and a Few Drops of Zeta Regularization*

It so happens that the present author, aside from being both a physicist and a mathematician, has dedicated a good part of his life to considering questions related to "infinity". Although, rather, only to some aspects of this concept, since "infinity" has many faces. So many, that more than one mathematician, like the great Georg Cantor, who dedicated a good part of his life to studying it, and decisively advanced his knowledge of it, in the end he went mad, literally [7,8].

Another famous mathematician, Leonard Euler, was convinced that all infinite series made sense; specifically, that with some reasonable criterion they could always be assigned a finite value [9,10].

To give a couple of examples of this fact, the simplest infinite series:

$$1 + 1 + 1 + 1 + 1 + 1 + \ldots \tag{1}$$

is better represented as corresponding to the value $-1/2$,

$$1 + 1 + 1 + 1 + 1 + 1 + \ldots \approx -1/2; \tag{2}$$

while the series

$$1 + 2 + 3 + 4 + 5 + 6 + 7 + \ldots \tag{3}$$

yields, with the same method (namely, analytic continuation by means of the zeta function), the value $-1/12$ [11–14]

$$1 + 2 + 3 + 4 + 5 + 6 + 7 + \ldots \approx -1/12. \tag{4}$$

and it happens that, when the most advanced physical theories make an appropriate use of such a discovery, this apparent "foolishness" or "arbitrariness" (once well worked out in the complex field [11–14]) acquires an extraordinary value in describing with incredible precision the phenomena of the real world.

Leonhard Euler devoted great efforts to studying infinite series. The harmonic series

$$1 + 1/2 + 1/3 + 1/4 + 1/5 + \ldots \tag{5}$$

fascinated him. Its terms correspond to the wavelengths of the successive notes of a vibrating string of a musical instrument—and each one of them is the harmonic mean of its neighboring terms. Mathematically, it turns out to be a logarithmically divergent series, but Euler realized that, by putting an exponent, *s*, on the natural number for the frequency

$$1 + 1/2^s + 1/3^s + 1/4^s + 1/5^s + \ldots = \sum_{n=1}^{\infty} \frac{1}{n^s} \tag{6}$$

he obtained for the sum of the infinite terms spectacular results, such as the following. When the exponent *s* was 2, the result was

$$\pi^2/6, \tag{7}$$

and for *s* = 4, 6, 8, he obtained, respectively,

$$\pi^4/90, \ \pi^6/945, \ \pi^8/9450. \tag{8}$$

Furthermore, Euler obtained all results up to the power 26. As a matter of fact, there is a closed expression for any even exponent, *n*, in terms of $\pi^n$ and Bernoulli numbers.

However, when the exponent is 1, we are back to the initial case of the harmonic series, and there is no way to obtain a finite value for it. This one is indeed a genuine logarithmic divergence; what means that, if we sum a number *N* of terms in the series, the value of the sum behaves as *log N*.

Euler derived important properties of this modified series as a function of *s*, not only for even or natural values of *s*, but also for other real values of this variable. One of the most important he obtained is the relation of this function to prime numbers: the zeta function is equal to an infinite product extended to all prime numbers, *p*, namely

$$\sum_{n=1}^{\infty} \frac{1}{n^s} = \prod_p \frac{1}{1 - p^s}. \tag{9}$$

Some years later, Bernhard Riemann went much further and discovered the extraordinary importance of the function of *s*, when *s* is allowed to be a complex number. He defined the zeta function, $\zeta(s)$—today called the Riemann zeta function—as follows:

$$\zeta(s) = \sum_{n=1}^{\infty} \frac{1}{n^s}, \text{ when Re } s > 1 \tag{10}$$

that is, when the real part of the variable *s* is bigger than *1*. There, the series is absolutely convergent, can be added without any problem and yields a finite result. However, what happens for the other values of *s* on the complex plane that do not satisfy this condition? Riemann showed that this series can be *uniquely* extended to all the rest of the complex plane, C. At any point, this analytic extension gives a finite value, with a single exception, that is, when *s* = *1*, in which case one recovers the harmonic series, inevitably divergent. This one, therefore, remains as the only singularity of the zeta function, in the entire complex plane, a simple pole with residue equal to *1* (Figure 3).

It is to be noted that the vertical line of the complex plane Re *s* = *1/2* (parallel to the ordinate axis) is of exceptional importance. Specifically, the Riemann conjecture (or Riemann hypothesis, considered by many to be the most important unsolved mathematical problem in history) states that the points in the complex plane where the zeta function is zero are all located on this line (one leaves apart, of course, the so-called trivial zeros, which are well known and obtained when $s = -2n$, for $n$ any arbitrary positive integer).

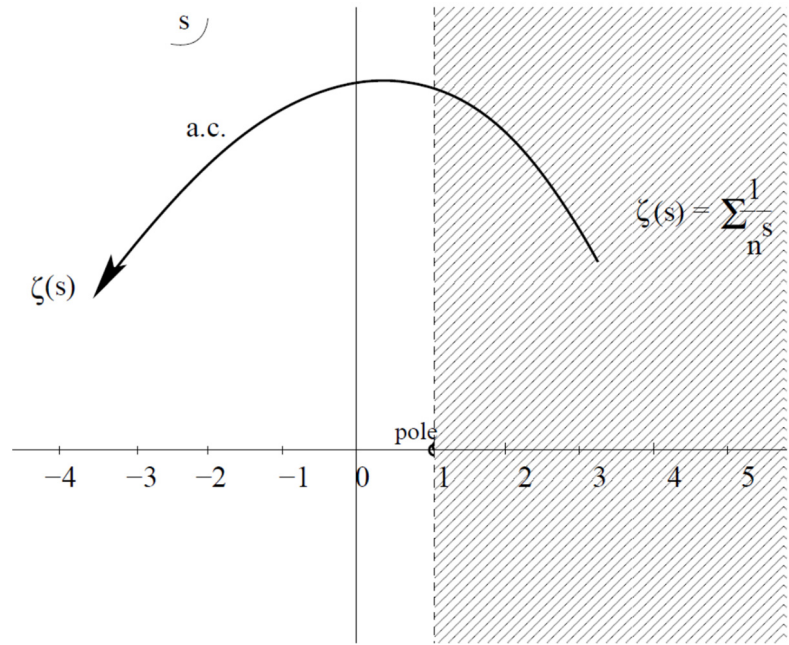

**Figure 3.** Definition of the Riemann zeta function on the complex plane.

The zeta function is a wonderful function and at the same time full of mystery. Many consider it the most important function ever constructed. We can look at her as an oracle, or a prophetess, and we shall see why. The author has devoted a significant part of his life to studying a few physical applications of it, and also of several of its non-trivial generalizations, as zeta functions build from the spectral values of pseudodifferential operators, which are associated with physical quantities [13,14] (one zeta function bears the author's name). Many references on these results can be found in [15].

In a few words, by means of the zeta function, terribly divergent infinite series can be driven to give finite values. What is more important is that, these values, which often may seem absurd at first sight, do make full sense in the end, in the physical world, having been validated by the most precise and accurate laboratory experiments ever conducted to date.

*2.4. A Couple of Simple Examples, and a Word of Warning*

If you look deeply into the matter, you will discover that all secrets of the zeta function are hidden in the extraordinary properties of the complex field. There, the most amazing of all mathematical formulas appears, namely

$$e^{i\pi} + 1 = 0. \tag{11}$$

It contains, in a single and very simple expression, the most important mathematical numbers. No matter how much one went on, it would be impossible to give even a faint idea of the exceptional importance that complex numbers have in physics and in all the sciences. Every scientist has a particular glimpse of this crucial importance, in his/her own field. Simply said, without them, no theory would be possible. Although, of course, they are not real numbers and, consequently, the result of the measurement of whatever quantity of nature will not be expressed under the form of a complex number.

We have already seen that in Riemann's definition of the zeta function these numbers play a key role, allowing the function to be defined in a unique (unambiguous) way, as a meromorphic function.

Let us now consider our first series (1) and assume that it is indeed a form (a particular case) of the zeta function. Actually, it is. So, for *s = 0*, that is

$$1 + 1 + 1 + 1 + 1 + 1 + \ldots \approx \zeta(0) = -1/2. \tag{12}$$

For the other case (3), we have similarly,

$$1 + 2 + 3 + 4 + 5 + 6 + 7 + \ldots \approx \zeta(-1) = -1/12. \tag{13}$$

For these are the values one obtains when one looks at the values of the zeta function at the positions *s = 0*, and *s = 1* of the complex field, respectively. To repeat, this may look at first sight as a mere mathematical trick, but it is indeed of enormous value because of its deep implications in the real world.

Nevertheless, dealing with these sums is not as simple as they might seem from the above. The present author has devoted a lot of time correcting the numerous mistakes in the mathematical–physics literature coming from erroneous applications of the method of zeta regularization, even though this actually is a rigorously defined procedure, which we cannot go into any detail here.

Let us just discuss a typical mistake that has been committed very often. Consider the series

$$\sum_{n=0}^{\infty} \left(n + \frac{1}{2}\right). \tag{14}$$

A common mistake, which could be made by anyone who thinks about what has been explained above, and who feel they have already become an expert in those calculations (this actually happens too often!), is to split it into 1/2 (corresponding to *n* = 0), plus two series, both starting from *n* = 1, as it corresponds to the zeta function definition, so that the result would then be

$$1/2 + \zeta(-1) + \frac{1}{2}\zeta(0) = 1/2 - 1/12 - 1/4 = 1/6. \tag{15}$$

This value is an error. The actual value is obtained by considering the whole series, as it is (*without* any splitting), which constitutes a form of an extended Riemann zeta function, namely the Hurwitz zeta function. At the value *s* = −1 it yields 1/24. The deviation (a factor of 4) could be crucial, if this would correspond to an actual physical experiment [12].

*2.5. Dealing with Divergences in Quantum Field Theory*

This brings to the author's mind conferences that two very competent theoretical physicists, Francisco Ynduráin and Andrei Slavnov, gave some years ago at the University of Barcelona. In them, both speakers invoked (independently from each other, on two different occasions) the above procedure and correspondences (in particular, Equation (1)) as the basis of the regularization methods they used. Both kindly suggested to the audience that anyone who did not know about them would do well to leave the classroom, since they would not understand any of the subsequent discussion. It was, frankly, amazing, as if they had copied their gestures from each other.

Such is the dependence of current quantum theories on the renormalization procedures of the infinities that so often appear in them. The story would be too long to tell at this point in any detail (see Section 2.6 for more information). Its origin can probably be found in investigations by Hendrik Lorentz and Max Abraham, who already at the beginning of the last century tried to develop a classical field theory for the electron, which eliminated the problems associated with infinities that arose by considering them as point particles.

Later, in the 1930s, the terrifying infinities that appeared in the perturbative series calculations of quantum field theories turned the greatest physicists of that time (including Paul Dirac and Albert Einstein, among many others) upside down. Robert Oppenheimer showed in 1930 that higher-order perturbative calculations in quantum electrodynamics (QED) always resulted in infinite quantities, such as the electron self-energy and the vacuum zero-point energy of the electron and photon fields [16,17]. Max Born, Werner Heisenberg, Pascual Jordan, and Paul Dirac, while developing quantum electrodynamics in the 1930s, also arrived at similar conclusions. In papers written between 1934 and 1938, a relativistic invariant formulation of quantum field theory (QFT) was established by Ernst Stueckelberg, who, by 1947, had developed a complete renormalization procedure (an achievement that was not properly recognized by the theoretical community of the epoch). It was not until some years later that a systematic and convincing approach to remove all the infinities was developed [18,19].

However, one had to wait for QED, which was the first successful case of a renormalizable quantum field theory. Renormalization was first developed for QED, in order to make sense of infinite integrals in perturbation theory. Initially viewed as a suspect provisional procedure, even by some of its originators, renormalization was eventually embraced as an important and self-consistent meaningful mechanism of scale physics in several other fields of physics and mathematics. This breakthrough came around 1950, when a robust method for eliminating infinities was developed by Freeman Dyson, Julian Schwinger, Richard Feynman, and Shinichiro Tomonaga. The last three received the Nobel Prize in Physics in 1965 for the quantization of electromagnetism, QED. All these works were later systematized by the first, Freeman Dyson.

The main idea is to replace the calculated values of the mass and charge, which are infinite in principle, by their finite measured values, in a fully consistent way. This systematic computational procedure is known as renormalization and can be applied to arbitrary order in perturbation theory. By using the renormalization procedure, calculations could finally explain the electron's anomalous magnetic moment (deviation from two of the electron *g*-factor) and also vacuum polarization. The results perfectly agreed with the experimental measurements, putting an end to the very harsh discussions, in which some reputed physicists had even predicted that those methods could lead to the end of physics.

Eventually, all the divergences could be renormalized by means of this precious theory: the theory of renormalization [16–19]. To have a physical feeling of what it is about, consider, as an example, a theory for an electron, which may start by postulating an electron with an initial mass and charge. However, it so happens that, in quantum field theory, a cloud of virtual particles, such as photons, positrons, and others, surround and interact with the initial electron. Accounting for the interactions of the surrounding particles (e.g., collisions at different energies) shows that the electron system behaves as if it had a different mass and charge from the ones initially postulated. Renormalization mathematically replaces the initially postulated mass and charge of an electron with the experimentally observed mass and charge, where these physical interactions have all been considered. Positrons and more massive particles such as protons exhibit exactly the same observed charge as the electron—even in the presence of much stronger interactions and more intense clouds of virtual particles. Renormalization specifies relationships between parameters in the theory when parameters describing large distance scales differ from parameters describing small distance scales. These relationships or transformations constitute a mathematical group, termed the renormalization group.

According to Wilson's picture, every QFT is fundamentally accompanied by its respective energy cut-off, $\Lambda$, with the meaning that the theory is no longer valid at energies higher than $\Lambda$, and that all degrees of freedom above the scale $\Lambda$ are to be omitted. For an example, the cut-off could be the inverse of the atomic spacing in a condensed matter system, and in elementary particle physics it could be associated with the fundamental "graininess" of spacetime caused by quantum fluctuations in gravity. The cut-off scale of theories of particle interactions lies far beyond current experiments. Even if the theory was very complicated at that scale, as long as its couplings are sufficiently weak, it must be described at low energies by a renormalizable effective field theory.

One difference between renormalizable and non-renormalizable theories is that the first are insensitive to details at high energies, whereas the last depend on them. According to this view, non-renormalizable theories are to be seen as low-energy effective theories of a more fundamental theory. The failure to remove the cut-off $\Lambda$ from calculations in a non-renormalizable theory merely indicates that new physical phenomena appear at scales above $\Lambda$, where a new theory, which we still do not have, is necessary. The understanding and the way to deal with effective field theories was decisively later improved by Steven Weinberg, as is mentioned elsewhere in the present article. Some consider this to have been his most important contribution to physics.

The idea of renormalization was later applied to other quantum field theories, described as gauge invariance theories. The last great finding was to show that such gauge field theories, corresponding to the weak and strong interactions, were also renormalizable. This was also recognized by the Nobel Prize in Physics, awarded to the discoverers of this fact, Gerardus 't Hooft and Martinus Veltman, in 1999 [18,19], an award with which several other physicists, who have applied the renormalization group to other very diverse disciplines, were also recognized, such as Kenneth G. Wilson. Further, the names of Nicolai Bogoliubov, Curtis C. Callan and Kurt Symanzik must also be mentioned in this context. For completeness, it must be added, too, that the regularization method for the infinities, used in this case and leading to the proof of renormalizability, was dimensional regularization, also an analytic continuation method, which was proven to work in all loop orders in perturbative calculation [18,19]. In that respect, the situation is not the same with zeta function regularization, which has a more limited scope. A still brief but more systematic presentation follows.

*2.6. Basics on Regularization and Renormalization*

As was already explained before, when divergences appear in a physical theory it is because the theory is ill defined, it is not valid in the domain or under the conditions that make divergences pop up, and must then be submitted to the following procedures.

2.6.1. Regularization

Popularly, the regularization process is viewed as a way to make divergent integrals or infinite sums convergent. This idea has been already discussed sufficiently well before.

A most important issue, to be stressed again, is that in this process a parameter is introduced, usually denoted with the Greek letter $\Lambda$ (or sometimes with $\mu$, but please do not confuse this $\Lambda$ with the cosmological constant), which plays a very important role. This parameter has a clear physical meaning, as a bound or cut-off marking the range of energies up to which our theory can be considered to be valid. One important example is the regularization of quantum vacuum fluctuations. In principle, there could be fluctuations at all energies or frequencies, but the fact is that no serious physicist will trust quantum field theory beyond the limits naturally imposed by the so-called Planck quantities (Planck's mass, Planck's length, Planck's time). As a consequence, in many circumstances those values are most reasonable choices for $\Lambda$.

Once such a parameter is introduced and the theory has been regularized, we can calculate any desired quantity in terms of the quantities appearing in the original Lagrangian (those we have started with, such as masses m, couplings $\lambda$, etc. so-called "bare" quantities) along with the newly introduced, regularization parameter, $\Lambda$. The bare quantities are not measured in experiments and are the ones that, in principle, will produce infinities. What is actually measured in experiments are the corresponding physical quantities (say, the physical masses $m_p$, couplings $\lambda_p$, etc.).

Regularization can be actually carried out by very different and well-stablished methods, which respond to the names: dimensional regularization, Pauli–Villars', Hadamard's, cut-off and lattice regularizations, zeta-function regularization, etc. They have proven to be incredibly successful in many different situations. Perhaps the most popular of them is dimensional regularization [20], which moves the integrals into a space with a (fictitious) fractional number of dimensions. Another is Pauli–Villars' [21], which adds fictitious particles to the theory with very large masses, so that loop integrands with the massive particles cancel out original loops at large momenta. Lattice regularization, introduced by Kenneth Wilson, replaces spacetime with a hyper-cubical lattice construct of fixed grid size, corresponding to a natural cutoff for the maximal momentum that a particle may possess when propagating on the lattice [22]. After doing a calculation on several lattices with different grid sizes, the result is extrapolated to grid size going to 0, in the hope that a scaling limit will exists and manifest itself in the series of numerical results, which will converge to the physical result.

Commonly, however, regularization only constitutes a first step towards the resolution of the main issue here, of assigning to the physical quantities of the theory the value that will precisely correspond to the one obtained in the physical experiment. A common mistake in the literature has been to take a finite result (product of the regularization) for the good, physical one we have been looking for.

It should be also noted that, for the specific case of zeta-function regularization, in a good number of the expressions to be found in the literature, the regularization parameter $\Lambda$ is missing (even if, actually, it is *always implicitly* there!). This comes from the fact that zeta function regularization is a procedure that stems, as we have just seen, from the mathematical summing up of divergent numerical series, where the regularization parameter, $\Lambda$, has no place. The numbers appearing in plain numerical series have, in principle, no physical dimension associated with them, and, at this purely mathematical level, the method simply reduces to an analytical continuation on the complex plane, in the way already explained in Section 2.3.

A very different case is when exactly the same idea is extended to a series that corresponds, typically, to the trace of the Hamiltonian of a physical system (e.g., the fluctuations of the quantum vacuum state corresponding to a given quantum field theory). In any situation of this sort, or similar, already at the very first stage of the proper definition of the zeta function corresponding to the physical operator, in this case the Hamiltonian of the system, a parameter, say $\Lambda$, with the physical dimension of mass (or energy) which must be introduced, in order that the zeta function corresponding to the operator can at all be defined (since the spectral values have to be rendered *dimensionless*, in order to work with them through mathematical expressions). This is, of course, a trivial reflection, but a very fundamental one in order to show that the regularization parameter is *always* there.

The fully-fledged method of zeta function regularization is schematically defined as follows. Take the Hamiltonian, H, corresponding to our quantum system, plus boundary conditions, plus a possible background field, and including a possibly non-trivial metric (when the effects of curved spacetime are important). In mathematical terms, all this boils down to a (generically second order, elliptic, pseudo-) differential operator, H, plus corresponding boundary conditions (see, e.g., [13,14]). The spectrum of this operator may or may not be explicitly obtainable and, in the first case, may or may not exhibit a beautiful regularity in terms of powers of natural numbers (to allow its manipulation with well-known zeta functions, as Riemann's, Hurwitz's, Epstein's, Barnes', Chowla-Selberg's, etc). Under quite basic conditions, a generalized zeta function, $\zeta_H$, corresponding to the operator H, can be (almost uniquely) defined in a rigorous way. A formal expression for this definition is:

$$\zeta_{H/\Lambda}(s) = \text{Tr } e^{-s \ln H/\Lambda}, \qquad (16)$$

which will make sense just in some domain of the complex plane s and only if H fulfills typical spectral conditions (as the Agmon-Nirenberg one). Let us stress that this definition is quite general, in the sense that it includes most interesting physical situations. In the particular case when the eigenvalues of the Hamiltonian (with the boundary conditions having been taken into account)—can be calculated explicitly (let us call them $\lambda_n$ and assume they form a discrete set, with n in general a multi-index, possibly with a continuous part), the expression of the zeta function is then given by:

$$\zeta_{H/\Lambda}(s) = \sum_n (\lambda_n/\Lambda)^{-s}, \qquad (17)$$

An expression valid to the rhs of the abscissa of convergence of the series. This abscissa is placed at the value obtained from the dimension of the manifold divided by the order of the operator, in a quite general situation, and reduces to Re $s$ = 1 (as we have already seen) in the case of the Riemann zeta function. That is, in all rigor, the starting setup of the zeta function regularization method that the interested reader may find in Refs. [13,14], explained in detail and with many different examples of its uses in physics.

It was Julian Schwinger who discovered a relationship between zeta function regularization and the renormalization he himself used in QED. In the meantime, other interesting connections among the different regularization methods have been found in the literature.

2.6.2. Renormalization

Renormalization is the process by which one takes the regularized theory, which is written in terms of the bare quantities and the regularization parameter (namely, $\Lambda$, m, $\lambda$, . . . ), and applies certain conditions, the renormalization conditions, which cause the physical quantities one wants to compute, such as energy levels or scattering amplitudes, to depend only on the physical quantities ($m_p$, $\lambda_p$, . . . ). When performing this procedure on a renormalizable quantum field theory, the dependence on the cutoff disappears.

In a sense, renormalization can be thought of as more of a procedure for writing the theory in terms of physical quantities than as a procedure for "removing infinities". The removal of infinities is already accomplished through regularization, but it takes renormalization to actually make sure that the finite values obtained by the regularization procedure do actually make physical sense, that is, that they do correspond to the ones we will obtain in a physical experiment.

As explained above, the regularization parameter, $\Lambda$, corresponds to a particular scale. The first step of the renormalization program is a systematic investigation of the changes of the physical system as viewed at different scales, when $\Lambda$ changes (or "runs"). A change in scale is called a scale transformation and these changes can be seen to constitute a group of transformations, called the renormalization group. In particle physics, it reflects the changes in the underlying force laws. The renormalization group is intimately related to scale invariance and conformal invariance, symmetries in which a system appears the same at all scales (self-similarity). In renormalizable theories, the system at one scale will generally be seen to consist of self-similar copies of itself, when viewed at smaller scales, with different parameters describing the components of the system.

Important contributions along this line in condensed matter physics were those of Leo P. Kadanoff, who in 1966 proposed the "block-spin" renormalization group [23], a way to define the components of the theory at large distances as aggregates of components at shorter distances. This conceptual approach was completed by Kenneth Wilson, who developed a new method in the theory of second-order phase transitions and critical phenomena, in 1971, and in 1975 obtained a constructive iterative renormalization solution of the Kondo problem. In parallel, in particle physics a formulation of the renormalization group equations was constructed by Curtis C. Callan Jr. and Kurt Symanzik in 1970 [24,25]. The so-called beta function, $\beta(g)$, which encodes the dependence of a coupling parameter, $g$, on the energy scale, $\Lambda$, of a given physical process in quantum field theory, is defined as

$$\beta(g) = \frac{\partial g}{\partial log(\Lambda)} \ . \tag{18}$$

Because of the existence of the renormalization group, $\beta$ has no explicit dependence on $\Lambda$, and just an implicit dependence through $g$. This dependence on the energy scale is known as the "running" of the coupling parameter. Moreover, it was found to amount to the canonical "trace anomaly", which represents the quantum-mechanical breaking of scale (dilation) symmetry in a field theory. Applications of the renormalization group to particle physics soon exploded with the establishment of the Standard Model.

In short, just briefly summarizing the key issue posed at the beginning and sufficiently discussed in the preceding subsections, the main conclusion to be drawn from all those developments is that, actually, there are no infinities in physics, in the real world. This constitutes an extremely valuable lesson. All the infinities that appear in any description or model of nature, expressed by means of mathematical equations, are always due to extrapolations, which are made beyond the range of validity of the equations used. They never correspond to physical reality, and they directly point out to the failure, the breakdown of the equations, in the region or under the conditions that originate the singularity.

*2.7. Further on the Big Bang Singularity*

Returning to the specific case of GR, the singularities that arise at the origin of the Universe, as well as inside each of the numerous black holes that exist in the Cosmos, are most probably not real. And this is so, as much as we read in the best encyclopedias and textbooks that such singularities are mathematically rigorous results, the outcome of flawless theorems, proved by the most brilliant minds, such as those of Stephen Hawking, Roger Penrose and others, and that have led to Nobel Prize awards. All the latter is absolutely true: they are in fact rigorous results obtained by exceptional geniuses and for the most brilliant of theories: GR. However, it turns out that, from the physical point of view, there is an Achilles' heel: as has been made clear before, this theory, GR, is not applicable, it is useless at very small scales; we enter there in a domain in which the GR ceases to be valid! It is no better than the Newtonian physics that, as everybody knows, had to be drastically modified at small scales and replaced by quantum physics.

Here, an important consideration is in order. In the popular literature, when talking about singularities, people have in mind, almost exclusively, those yielding horrible infinities for the most usual physical quantities: energies, masses, charges, etc. However, it turns out that the singularities appearing in the famous theorems of Penrose, Hawking, and others related to them, are of a very different kind. Those are, in fact, incompleteness theorems, in relation with geodesics, idealized particle paths moving towards the past (the singularity at the origin of the Universe) or towards the future in time (black hole singularities). Those are not actual singularities, in the sense that some physical quantity explodes. It is just that time simply disappears, it ceases to exist, it is not there before or after some point in the geodesic path. This is an important difference to be considered when talking of singularities. However, the conceptual problem associated with the fact that the concept of "time" itself has no sense, in any known physical theory, beyond the bound established by Planck's time, still persists.

And, on top of it, the above-mentioned singularity theorems do not take into account two very important circumstances concerning the evolution of the Universe. The first issue is, namely, that inflationary models seemed to invalidate the conditions of Penrose's and Hawking's theorems. This was eventually proven not to be the case, by A. Borde and A. Vilenkin, in a paper from 1994, where they showed that inflationary universes are still geodesically incomplete towards the past [26]. This was further extended in a theorem by A. Borde, A.H. Guth and A. Vilenkin, who in 2003 proved that the essential conclusion of past geodesic incompleteness is still maintained under the conditions that are compulsory to obtain chaotic inflation [27].

However, there is more to this. Namely, it has actually been possible to demonstrate that sometimes it is enough to correct the equations of GR just a little, to patch the theory up by adding a pinch of quantum physics to the classical GR formulas (in technical language: first-order quantum corrections) and the infinities suddenly disappear, like magic, in a similar way to how the singularities of classical electrodynamics vanished when the theory was quantized, giving rise to quantum electrodynamics (Feynman-Schwinger-Tomonaga, see above). However, in gravitation it has not yet been possible to complete this process thoroughly, nor is there any promising glimpse of how this could be done: this was one of the objectives of supergravity, which still remains pending.

The question is now: were the great scientists who proposed these rigorous singularity theorems aware of the limitations just mentioned? The answer is yes, of course. One only has to go to the website of the late Stephen Hawking to see everything that has been argued here masterfully explained. Hawking was precisely one of the first to add the quantum corrections that have been mentioned, after a visit he made to the Soviet Union in 1973, where he discussed—as he himself confessed in his memoirs—with Yakov B. Zeldovich (who did the first calculation of the value of the cosmological constant and interpreted it as coming from quantum physics) and his students in Moscow, pioneers in this field— corrections that, for the case of a black hole, led him to find the famous "Hawking effect", the most important discovery of his life [28,29].

However, this "tagline" of the theorems, that has to do with the limitations of GR, is explained in very few places because—despite its enormous significance—for the purpose of the apotheotic narrative usually pursued, it would be like the rain that puts out fireworks: all singularities and infinities would suddenly vanish. Again, there is more. Taking such quantum corrections to Einstein's field equations very seriously, and extending them to higher orders, theorists have built families of what are known as modified gravity theories, in particular the $f(R)$ theories, tensor–vector theories, and others, which are very promising, with a view to being able to solve some of the fundamental problems around, such as the nature of dark energy (and the cosmological constant issue) and dark matter. Several of these theories have been proven to be able to provide a unified cosmic history within the same given model, and also to yield workable answers to the dark energy and dark matter problems [30,31].

Before going into that, let us close the present section with the earnest request that, from now on, no one else be fooled again by all these "half-hearted" stories, unprofessionally told, with infinities and singularities galore, that only seek what the French call "*épater le bourgeois*". It seems that, today, if there is not a good dose of spectacularism in what is being said, there is no chronicle or article, because it does not sell.

## 3. The Real Big Bang: Inflation

From now on, let us focus on what is actually known, either by very clear evidence from laboratory experiments or through increasingly precise observations of celestial objects. To begin with, it is known that our universe was born some 13.78 billion years ago, since, on one side, no matter how carefully observations of the sky have been made, no older objects have ever been found. This clearly indicates that our universe was born then. However, there is also further independent evidence of this fact. The evidence is collected in the working Big Bang cosmological model.

It is also rigorously known that, shortly after its origin, when the Universe was between a trillionth and a millionth of a second of age, it was all a very hot plasma, made up of leptons, quarks and gluons. This is known as the primordial plasma, or as the quark-gluon plasma, or lepton-quark-gluon plasma. Leptons are the lightest elementary particles, quarks are the simplest components of non-leptonic elementary particles, and gluons are the interaction forces that hold them together, to form mesons (particles of intermediate mass, made up of two quarks), and baryons (the heaviest particles, consisting of three quarks). The properties of the primordial plasma have been very well determined in laboratories such as CERN in Geneva, Switzerland. However, how the hell did that plasma come to be? And from what preexisting stuff?

There is no definitive answer to those questions, but evidence is accumulating, which indicates that it was as a consequence of a sudden and very powerful expansion of the fabric of space, which enormously swelled it, as if it actually were a balloon (whose radius would correspond to time). Rigorously, the simplest model implementing this view would be a three-dimensional surface of a four-dimensional sphere, with the fourth (radial) dimension being time (multiplied by the speed of light squared). This is roughly what the Big Bang cosmological theory says. By the way, one should never confuse it with the Big Bang singularity [1,2], which has become nowadays the main common meaning of the tag "Big Bang". Neither of them reflects to the true, original meaning of this term, which is namely what is now known as "cosmic inflation" (see, e.g., Ref. [1]).

The evidence supporting the BB theory is very extensive and precise. The thermal fingerprint of the primordial plasma, the so-called cosmic microwave background (CMB) radiation that permeates the Universe, has been discovered and carefully studied. In a few words, this theory is a very successful one, as it also explains the current composition of the cosmos, as well as many other phenomena. However, it turns out that a few years ago, just before the end of the last century, it was discovered that the Universe is in fact expanding at an ever-increasing rate, in other words, the expansion is accelerating. This has generated a great mystery that the Big Bang theory fails to explain convincingly. Some theorists relate it to the accelerated expansion of the early cosmos, and modified gravity models have been built (in particular, those of the $f(R)$ type already mentioned above), which promise to provide a *unified* description of its entire history, corroborated by the data obtained from the most precise astronomical observations. Be that as it may, we still keep asking ourselves: What exactly happened when our Universe was born? And how did it take the form that it now has?

There is currently no valid physical theory to describe the origin of the Universe and only a few plausible proposals. The most advanced scientific ideas that are in no contradiction with known physics—but not yet tested in the laboratory, either—allow us to go back to when the cosmos was only $10^{-35}$ or $10^{-36}$ seconds of age. That is, up to a trillionth of a trillionth of a second, after everything started.

At that time, the Universe underwent an extremely brief and dramatic stage of inflation, expanding much faster than the speed of light. It doubled in size probably as much as one hundred times, in a tiny fraction of a second, and its temperature dropped from some $10^{27}$ to some $10^{22}$ K.

Here, a comment is in order. It might seem that inflation would violate the main principle of the special theory of relativity, but actually, it does not. In fact, relativity holds that neither information nor matter can be transported between two positions in space at a speed faster than that of light. However, inflation did not involve any space travel, as it was just an expansion of the fabric of space itself.

Before inflation, there was hardly anything: just a handful of ingredients—only those strictly necessary to set everything in motion are needed: space-time itself, the Higgs field, the field that receives the name of *inflaton*, to spark inflation, and little else. As of today, there is no proven scientific evidence on how these ingredients appeared. However, what does seem quite clear is that it took something like inflation to suddenly enlarge the Universe and thus endow it with the properties we now observe. It should be mentioned, however, that there are alternative approaches that can dispense with the whole inflationary paradigm, completely replacing the image of the Universe, as a whole, with a new one, such as those that consider our Universe to be inside a black hole solution of Einstein's equations [32–35].

Such a rapidly expanding universe was practically devoid of matter, but, according to the most accepted theory, it harbored large amounts of dark energy. This is precisely the same force that is also believed to be driving the current accelerated expansion of the cosmos.

*3.1. Reheating*

Dark energy caused inflation and when, after a very brief moment in time, it stopped, the energy it harbored was transformed into ordinary matter and radiation, through a process called reheating. The Universe, which had been very cold during inflation, heated up to a very high temperature again (about $10^{25}$ K, according to some models) and the created matter was constituted in a primordial plasma, made up of leptons, quarks, and gluons, which are the most elementary components of all the matter we see now. However, there is no clear scientific evidence yet on how and why inflation appeared. Alternative mechanisms are being considered as well.

*3.2. Inflation or a Cyclic Universe?*

Inflation is, today, the most appropriate theory to explain the current characteristics of the Universe; in particular, the fact that it is, on large scales, so flat, homogeneous, and isotropic, with more or less the same amount of matter distributed equally at all its points and in all directions. Several pieces of evidence point to the fact that inflation did occur, but there is as yet no conclusive proof that this was the case.

Consequently, other alternative theoretical models cannot be ruled out, such as the so-called cyclical model, which recalls the Stoic philosophers' concept of the ekpyrotic universe. This argues that the Universe did not appear from a point, out of nowhere, but rather bounced from another pre-existing universe into expansion, at a much calmer rate than inflation theory predicts, from a previous universe that would have been cooling and then contracting. If that theory were correct, our universe would have undergone an indeterminate succession of previous expansions and contractions.

The cyclic model relies on string and brane theories. In accordance with them, it posits that the Universe actually has eleven dimensions, only four of which can be observed (the so-called space–time brane). Branes actually exist in more dimensions than the conventional four, being that the brane is the whole structure, and there are theories with different space–time dimensions. There could be other branes lurking in eleven-dimensional space, and a collision between two branes could have rocked the cosmos, shifting it from contraction to expansion and stimulating the CMB radiation that we now detect, and which would closely match that of the Big Bang model.

**4. The Universe We Know Takes Shape**

Just after inflation, the Universe was a very dense, absolutely dark, very hot plasma. When it was about a trillionth of a second ($10^{-12}$ s) young, it entered the electroweak phase, in which massless vector and scalar bosons were present. Shortly after, the photons were formed, from the first ones, and the Higgs boson acquired mass to, subsequently, provide it to the rest of the bosons; all this in perfect agreement with the electroweak quantum field theory [16,17]. The Universe was then about 300 light-seconds across (comparable to the distance between the Sun and Venus).

As time passed, the cosmos continued to cool, until it entered the stage, already mentioned above, in which quarks and gluons were formed. When the Cosmos was already a microsecond of age ($10^{-6}$ s), from the quarks and gluons, the first protons and neutrons could be constituted and, to an even greater extent, also the first mesons (muons and pions, fundamentally). In addition to their corresponding antiparticles, their production was soon halted by an asymmetry, providential for our existence and that of the Universe as we now know it, namely the baryon asymmetry, which has not yet been fully explained, but which is most crucial.

All the above has been reproduced very faithfully, step by step, in high-energy laboratories, such as the LHC at CERN, Geneva (which, in fact, can reach up to $10^{-15}$ s of age of the cosmos), and also in large heavy-ion colliders. When the Cosmos had one second of existence, the neutrinos managed to free themselves from the original soup, and they could already travel everywhere through the Universe. This was then about ten light-years across (a couple of times greater than the distance to the closest stars to the Sun).

After the first three minutes [36], protons and neutrons began to merge, forming deuterium nuclei; and these, on its turn, joined with each other, to form helium-4 nuclei.

*4.1. Recombination: The Universe Becomes Transparent*

Atomic nuclei could not yet constitute atoms, since the Universe was too hot to allow the capture of electrons by the protons. The first atom that could be formed was hydrogen, the simplest atom of all (one proton - one electron). This happened when the Universe was 378,000 years old, a period that is known as recombination. In a relatively short time span, the hydrogen ions, and also those of helium, were able to capture electrons, thus forming electrically neutral atoms, which, as we know, are extraordinarily transparent to light. And thus, of a sudden, the photons were finally free to cross the entire Universe. For the first time ever, there was light, which definitively separated from the darkness that had reigned until then (Figure 4).

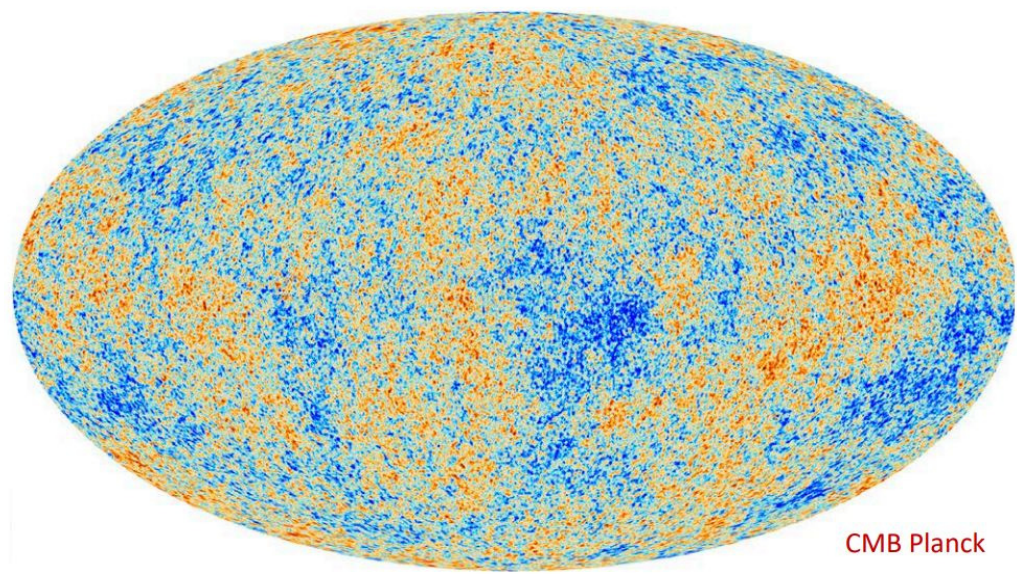

**Figure 4.** The cosmic microwave background radiation as reported by Planck.

Recombination radically changed the appearance of the Universe, which until that period had been very dense and opaque, a dark mist, suddenly becoming transparent. The cosmic microwave background radiation (CMB) that we continue observing today, although very faintly, is the light of that time.

*4.2. Reionization*

However, the Universe, as a whole, was still pretty dark for a long time after recombination. The expansion was constantly enlarging it, and there were still no objects of great luminosity around, apart from the faint background radiation. The Cosmos only truly lit up when the first young stars began to shine, some 300 million years after its birth. The early stars, and perhaps some other sources as yet little-known, or still to be discovered (remote objects that no one had counted on previously, are continually turning up in cosmos research), blasted enough radiation into Space to separate, again, most of the hydrogen atoms present in its constituent protons and electrons. Earendel is one of the most distant stars that have been observed from that time (Figure 5).

This process, known as reionization, appears to have run its course at least a billion years after the Big Bang. However, keep in mind that now the Universe was no longer opaque, as it had been before the recombination, since it had already greatly expanded, and its density was much lower. Today, photon scattering interactions are in fact quite rare, on a large scale, although they remain of paramount importance on local scales within large galaxies.

Over time, and by the action of gravity, the stars formed larger and larger galaxies and cosmic structures, interconnected on a large scale; it is what is called the cosmic web (cosmic web, Figure 6). The planets were later formed around some newly formed stars, as was the case with our Sun, and 3.8 billion years ago, life took root on Earth. However, this would already be a separate chapter, just as or even more exciting than the present one, as was said at the beginning.

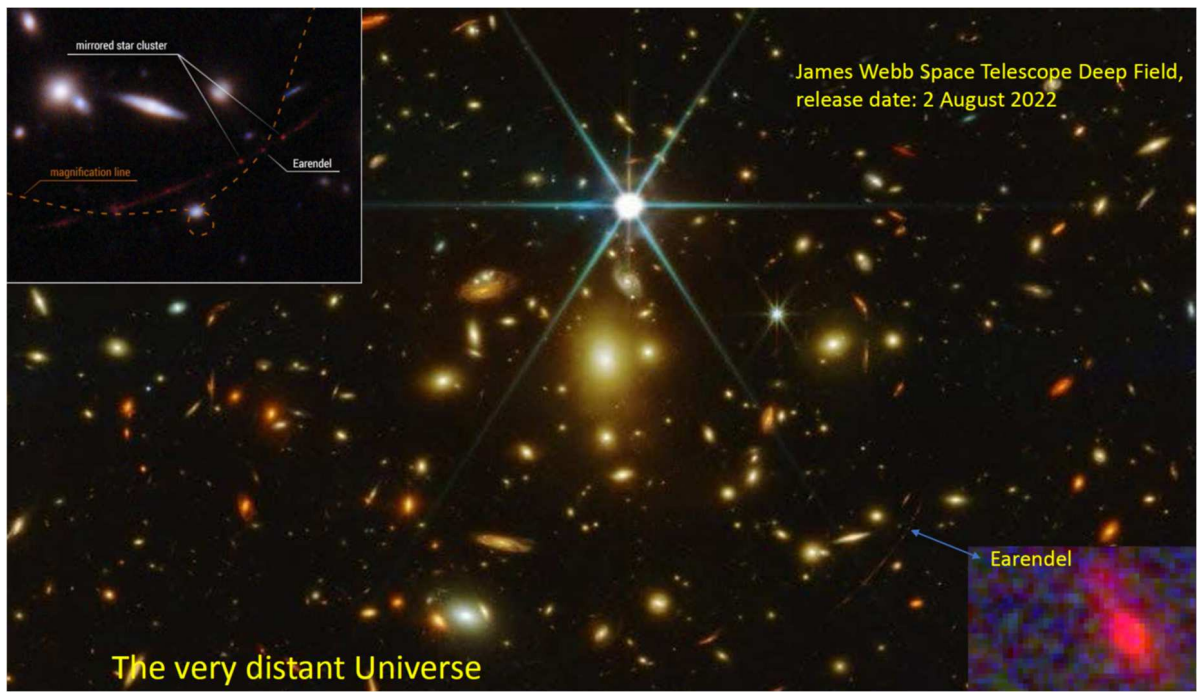

**Figure 5.** The very distant Universe. James Webb Space Telescope Deep Field, release date: 2 August 2022. Earendel is to be seen at the crossing of two gravitational lenses images.

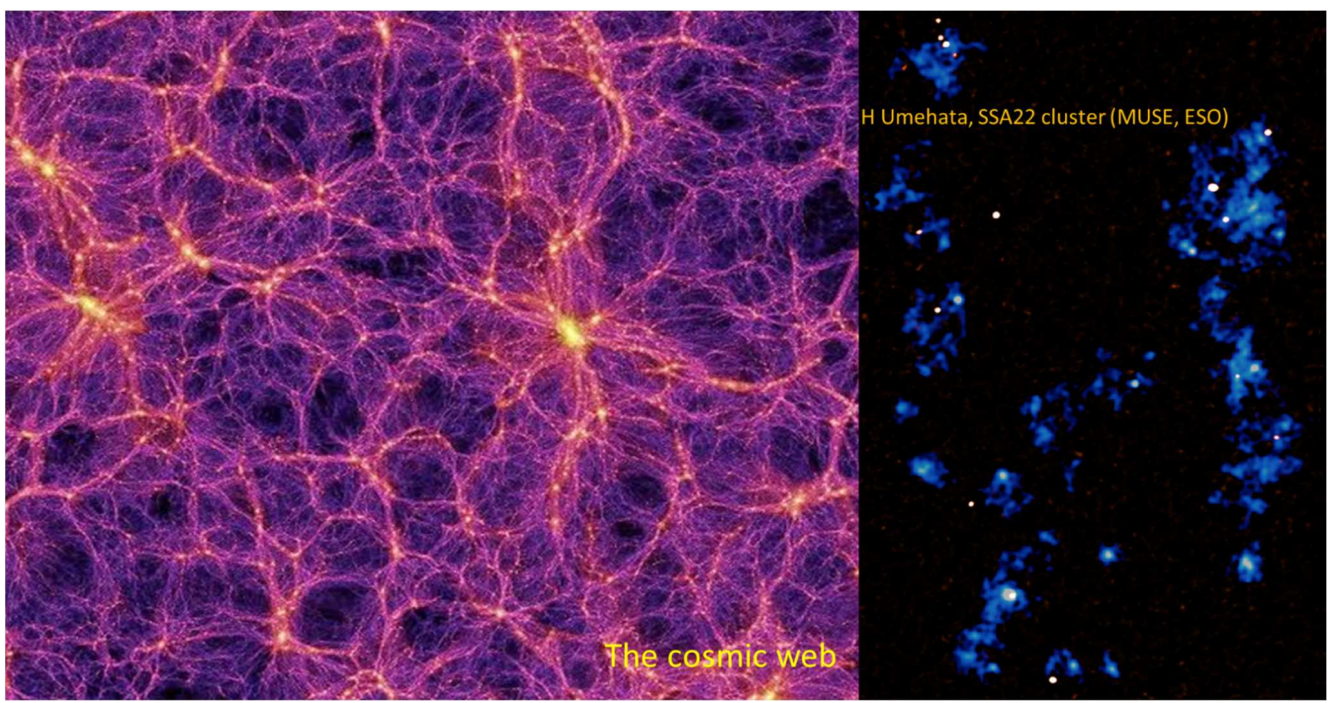

**Figure 6.** A comparison of the cosmic web as generated by computer simulations with a real observation by H. Humehata of the SSA22 cluster (MUSE, ESO).

## 5. Some Current and Future Missions

The Planck mission of the European Space Agency (ESA), which orbited the Earth between 2009 and 2013 (Figure 7), helped further refine theories about the nature of the cosmos and its origin [37,38]. The detailed map of the cosmic microwave background, generated by the mission, clearly revealed that our universe—although it cannot yet be completely ruled out that it might have arisen from a predecessor—will most likely never shrink again in the future, and this goes against the possibility of a cyclic universe.

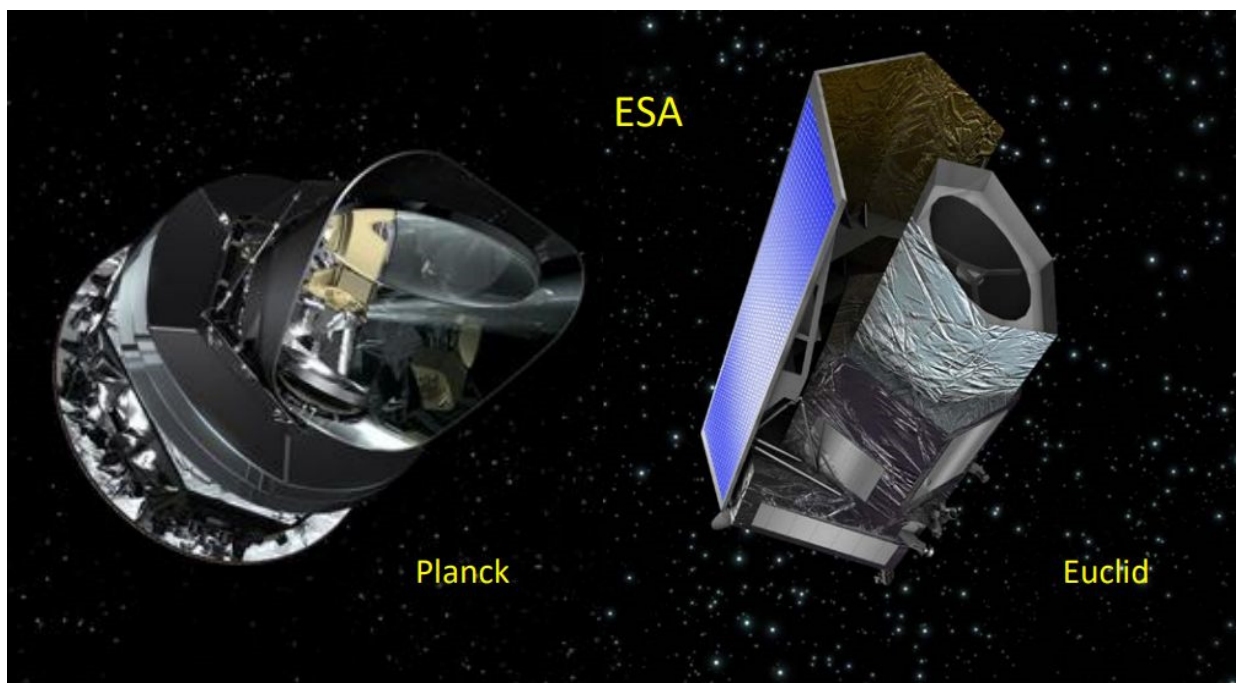

**Figure 7.** Artistic representation of the two ESA missions Planck and Euclid put together for visual comparison.

Using the Planck data, estimates of the age of the Universe and of the amounts of visible matter, dark matter, and dark energy it contains have been refined. The mission did not in fact provide any great surprises, largely confirming existing theories and corroborating that the future evolution of the Universe will be, in short, quite sad, or boring.

However, their results have also raised new questions. For example, the Hubble constant—which describes the rate of expansion of the cosmos—measured by Planck in the distant Universe, gives a different value than that obtained with the Hubble Space Telescope in the nearby Universe. This important tension between the two outcomes is now being feverishly investigated.

From the field of theory, while returning now to the infinities, with which we began this journey—and also answering questions that are often formulated by the audience when presenting these results—it must be added that modified gravity theories, such as those of the $f(R)$ type, although they are capable (as pointed out already) of eliminating singularities that appear in GR, they often create new ones, both in the past and in the future of the Universe. Thus, they do not constitute, in fact, any panacea in that sense. This warns us, being consistent, that they cannot, in any case, be final theories to describe the Cosmos. However, one should be aware, too, that not even a future theory of quantum gravity, nor probably the use of string theories, will literally be able to bridge the temporary distance between t = 0 and t = $10^{-44}$ s, the Planck time. Although it is most likely that, in fact, being able to go beyond those t = $10^{-44}$ seconds might completely lack any physical sense, if it can be shown that quantum physics is, de facto, the last (and, as such, insurmountable) frontier of physics. There are some theorists working in that direction.

As far as observations of the Cosmos are concerned, it is expected that ESA's Euclid mission, which will be launched in 2023 (Figure 7) and, a decade later, in the 2030's, the Einstein Telescope of gravitational waves (ET Project, Figure 8), the LISA mission, as well as a handful other projects, will help us advance in an understanding of the nature of the dark side of the Universe, namely dark energy, dark matter, black holes, as well as in resolving the other great unknowns already mentioned, and which still remain open for a comprehensive knowledge of the Cosmos [39–46].

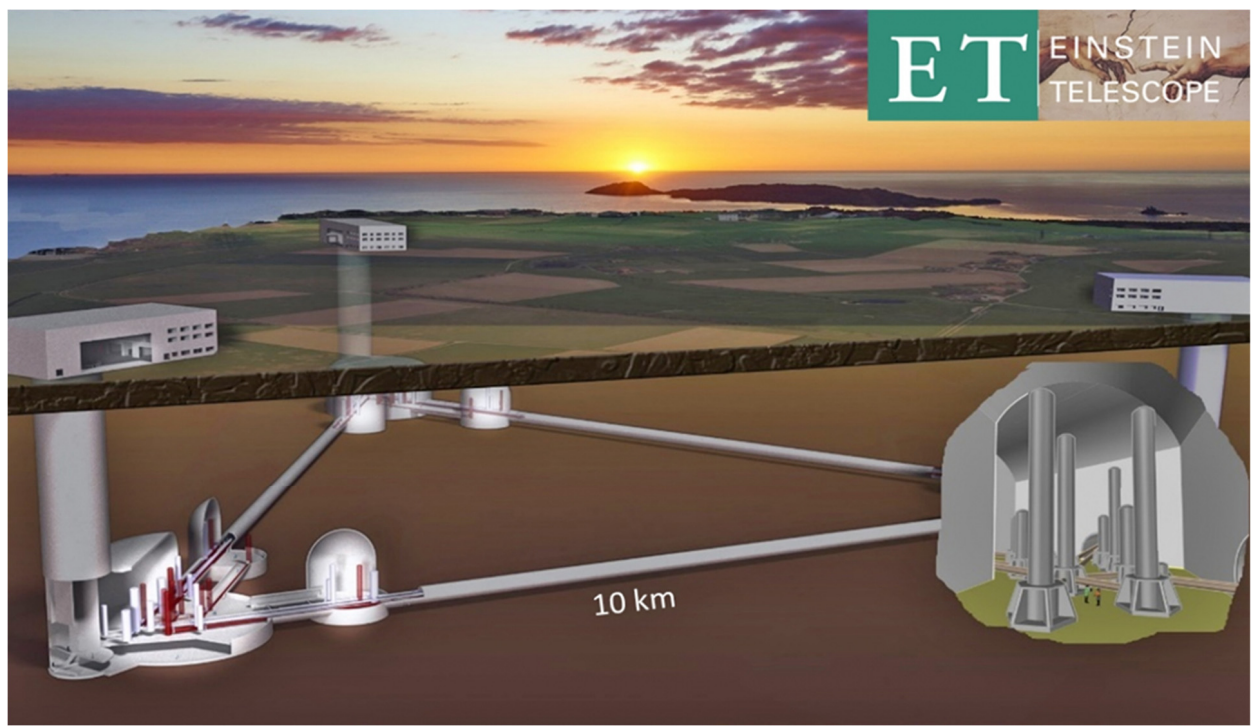

**Figure 8.** Schematic representation of the Einstein Telescope set up. The final site where it is going to be built has not been chosen yet.

## 6. Discussion

The high status of Einstein's General Theory of Relativity as the fundamental theory for explaining, in precise mathematical language, the origin and the behavior of the Universe, is still unchallenged. This theory, conceived by the author as a model for the rather small, static, and eternal universe of his time—which it was thought to just comprise the Milky Way—has now proven able to deal with a large and very diverse number of unforeseen scenarios playing in a million times larger cosmos, which is expanding at an accelerated pace. This has brought, as a consequence, the idea that GR is a kind of all-powerful theory and that we must believe wholeheartedly all the results and outcomes from it, whatever the situation and circumstances. This is to be compared, in a way, with what happened with the Newtonian universal gravity laws, in the course of several centuries. This is true, especially, for the abundant popular accounts of the Universe, which have proliferated since the Hubble telescope started to send spectacular images of galaxies and of other celestial objects [1], making of the observation of the Cosmos a thrilling social endeavor.

However, we should recall (although it may seem unnecessary, for it is rather well known) that, notably, Einstein himself was the first to point out—while constructing his theory more than one-hundred years ago—that his GR was in fact a truncated theory, at most a good approximation to a really fundamental one that would incorporate all of the physical ideas of his time—e.g., Mach's principle and the like—a variant of which has been called by some total relativity. Following Einstein, many other relevant physicists have since stressed the shortcomings of GR for dealing, e.g., with very large curvatures, energies, densities, or pressures [1,47]. Several of those situations have started to become crucial recently in different astrophysical observations. Some of them must, and can (at least to some extent) be dealt with QFTs of gauge interactions—in a word, the standard model of particle physics, which constitutes, so to say, the other (and already unified) side of the present-day fundamental description of nature and the Universe.

It was Franck Wilczek who, in an obituary for Steven Weinberg, stressed the importance of the work of this great physicist by saying that "he played a central part in formulating and establishing theoretical physics' two standard models—the standard model of fundamental interactions and the standard model of cosmology" [48]. The most important, and still missing, task to be fulfilled is to merge both together by building a model of quantum gravity. That seems, at the moment, impossible by the usual means. Now, my point here is that, outside the community of specialists, many scientists and a vast majority of science popularizers seem to forget about this double-sided approach to nature and about the necessity to match both sides.

Some paragraphs have been devoted, in the paper, to discussing these issues, in particular concerning the Big Bang singularity, with hints towards its possible disappearing when modified gravity theories are introduced to the game. We are far from having a reliable understanding of the origin of the Universe but, rather than spending much time on that (what is done all too often, not being of any practical use), it has been argued above that it is much wiser to spend it in obtaining a better knowledge of what is really known, which starts just a tiny fraction of a second after the singularity at t = 0.

Humankind should be extremely proud of this precious knowledge, accumulated over many centuries of constant, tireless research: that is, to start with, the fact that we human beings have been able to explain, with the help of very simple models, the behavior of the Universe we live in. It is this positive view which should be conveyed to students and to popular audiences, instead of spending most of the time on the many questions to which we do not now have answers. It is in this sense that a vision of the state of the art concerning the epoch, very close to the Universe's origin—which can be carefully investigated at the laboratory—and its subsequent evolution (more and more data is becoming available from space missions and terrestrial observatories), have been discussed.

On top of that, back to the issue of the singularities, some specific points have been picked up and conveniently highlighted: (i) The fact that all fundamental theories of physics (GR included) have a *limited* range of applicability; (ii) That, when extrapolated *beyond* their domain or any other conditions of validity, the theories produce singularities; and (iii) That, vice-versa, when one encounters a singularity, in a physical situation, it is always because the mathematical equations employed to implement the physical law or principle are simply *not valid* in the region where the singularity occurs.

Needless to say, singularities are most useful, and they have very often been along the history of physics; e.g., as indications that the theory at hand must be improved or sometimes completely changed in the circumstances that lead to the singularity. Several important examples of this fact have been provided above, and also some particular ways as to how the infinities can be, and have actually been, dealt with in the past already.

Finally, some terrestrial projects and space missions yet to be launched have been highlighted. There is a general conviction that many surprises are waiting for us in our present explorations of the Cosmos. Several of them are even expected to be already there, in the big files of observational data that have been taken already, and which are still waiting to be processed by data handling facilities. Indeed, a good number of celestial objects have already been found that defy our present models of the Universe and the standard descriptions—both at the GR level and also at that of the QFTs—starting with the well-known problems of the missing explanations for the nature of its dark side, namely dark matter, dark energy and black holes, but notably also concerning the possibility of quark stars having been already detected, as further alternatives to the omnipresent neutron stars and leading to a new paradigm of matter compression (with a, still unknown, new equation of state) [49,50]. Neutron stars and black holes with strange masses and properties have been found that defy the corresponding standard wisdom as well.

We are constrained to explaining all these phenomena with the laws of the standard models at disposal, although some alternatives have been proposed already, as different families of the so-called modified gravity theories and other kind of alternatives to GR; and also, some extensions of the standard model of particle physics, leaving room for pen-taquarks and other exotic phenomena [51,52]. Concerning supersymmetry and superstring theories, not many impacting news have been heard recently, while waiting for specific results from CERN or elsewhere, but they are of course latent and very well-grounded possibilities whose existence—in spite of the lack yet of actual experimental evidence—we should always keep in mind.

## 7. Conclusions

To put it simply, by means of the zeta function, awfully divergent infinite series can be brought to produce finite values. The method, which started being simply a convenient procedure to "give sense" (Euler) to divergent mathematical series, has been dramatically extended, both in mathematics—where it is key in the definition of divergent traces and determinants, e.g., in the celebrated Atiyah–Singer theory—and in physics, as a rigorous and widely used regularization procedure. What it is more important, the obtained results, which often may seem absurd, at first, do make full sense in the end, in the physical world, having been fully validated by the most accurate laboratory experiments ever conducted to date.

An important lesson to be taken home is that all fundamental theories of physics (GR included) have a limited range of applicability. When extrapolated beyond their respective working domains or any other conditions of validity, the theories tend to produce singularities. Vice versa, when one encounters a singularity in a physical situation, it is because the mathematical equations employed to implement the physical law or principle are plainly not valid in the region where the singularity pops up.

The moral: from now on, no one else should be fooled by all these abundant "half-hearted" stories, some of them unprofessionally told, with infinities and singularities galore, and which only seek to, what the French call, "*épater le bourgeois*". It seems that, today, if there is not a good dose of spectacularism in what is being said, there is no chronicle or article, because it does not sell. Needless to say, I am here only speaking of some bad popular articles, and not at all of those written in serious professional journals or by competent popularizers of science, of which, fortunately, there are many. Not in vain, this paper emerged from some popular talks, which I feel it is the duty of all professionals to deliver, at least from time to time.

**Funding:** This research was funded by MICINN (Spain), project PID2019-104397GB-I00, of the Spanish State Research Agency program AEI/10.13039/501100011033, by the Catalan Government, AGAUR project 2017-SGR-247, and by the program Unidad de Excelencia María de Maeztu CEX2020-001058-M.

**Data Availability Statement:** Not applicable.

**Acknowledgments:** The comments of the referees are gratefully acknowledged, since they have contributed to an improvement of the final version of the paper.

**Conflicts of Interest:** The author declares no conflict of interest. The funders had no role in the design of the study; in the collection, analyses, or interpretation of data; in the writing of the manuscript; or in the decision to publish the results.